\documentclass[aps,prl,superscriptaddress,twocolumn]{revtex4}
\usepackage{bm} \usepackage{graphicx} \usepackage{amsmath}
\usepackage{amssymb} 

\begin{document}

\title{Matter Wave Interferometry of a Levitated Thermal Nano-Oscillator Induced and Probed by a Spin}

\author{M. Scala}
\affiliation{Department of Physics and Astronomy, University College
London, Gower St., London WC1E 6BT, United Kingdom}

\author{M. S. Kim}
\affiliation{QOLS, Blackett Laboratory, Imperial College London, London SW7 2BW, United Kingdom}

\author{G. W. Morley}
\affiliation{Department of Physics, University of Warwick, Gibbet Hill Road, Coventry CV4 7AL, United Kingdom}

\author{P. F. Barker}
\affiliation{Department of Physics and Astronomy, University College
London, Gower St., London WC1E 6BT, United Kingdom}

\author{S. Bose}
\affiliation{Department of Physics and Astronomy, University College
London, Gower St., London WC1E 6BT, United Kingdom}

\date{\today}

\begin{abstract}
We show how the interference between spatially separated states of
the center of mass (COM) of a mesoscopic harmonic oscillator can be
evidenced by coupling it to a spin and performing solely spin
manipulations and measurements (Ramsey Interferometry). We propose
to use an optically levitated diamond bead containing an NV center
spin.  The nano-scale size of the bead makes the motional
decoherence due to levitation negligible. The form of the
spin-motion coupling ensures that the scheme works for
thermal states so that moderate feedback cooling suffices. No
separate control or observation of the COM state is
required and thereby one dispenses with cavities, spatially
resolved detection and low mass-dispersion ensembles. The
controllable relative phase in the Ramsey interferometry stems
from a gravitational potential difference so that it uniquely
evidences coherence between states which involve the whole
nano-crystal being in spatially distinct locations.
\end{abstract}

\maketitle

Quantum mechanics does not itself provide any limits to its
applicability. Its formalism is independent of mass -- an isolated object, however massive,
should exhibit superpositions. While it is important to test
whether this is true for a mesoscopic object, such tests face the
obstacle of decoherence. Decoherence, caused by the coupling of a
system to its environment, causes a reduction in the visibility of
interference patterns that evidence superpositions
\cite{Leggett,Zurek}.  Fundamental modifications of quantum
behavior may add to this \cite{GRWP,Penrose,RMP}. The pragmatic
approach would be to better isolate mesoscopic objects and
engineer quantum behavior. This has been happening successfully --
a superconducting system originally invoked to probe the
limitations of quantum behavior \cite{Leggett-Garg}, is now used
as a qubit \cite{flux-qubits}. Matter wave interferometry
with macromolecules has extended the quantum realm
\cite{Arndt}.  Further extension may be possible by coupling nano-cantilevers to photons
\cite{Bose-Jacobs-Knight,Marshall}, or qubits
\cite{Armour-Blencowe-Schwab,Rabl}. Optical counterparts
have also been explored \cite{Thew}. A nano-scale object levitated
by a (classical) light field can be fruitful for demonstrations of
quantum behavior \cite{Barker,Romero-Isart,Chang}.

\begin{figure}
\begin{center}
\includegraphics[width=2.0in, clip]{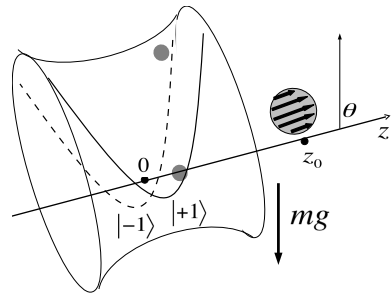}
 \caption{Setup: An optical trap holds a diamond bead with an NV center with both weakest confinement and spin quantization along the z axis.
 A magnetized sphere at $z_0$ produces
 spin-dependent shifts to the center of the harmonic well. An angle
 $\theta$ between the vertical and the $z$ axis places the centers of the wells corresponding to the $|+1\rangle$ and $|-1\rangle$ states in
 different gravitational potentials. A random coherent state of the center of mass of the bead oscillates as different coherent states in the two
 wells (grey filled circles), accumulating a relative gravitational phase difference due to superpositions. At
 $t_0=2\pi/\omega_z$ this phase can be read from spin measurements.}
\label{setup}
\end{center}
\end{figure}

In this letter we first enlist five desiderata which would simplify
the path to extend quantum mechanics to larger objects and then
propose a Ramsey interferometry experiment on a spin-mechanical
oscillator hybrid system that meets these desiderata. Firstly, it
is challenging to create nearly ``pure" non-classical states of
the center of mass (COM) of mesoscopic objects, requiring cooling of the
object to its ground state, though there is much work in this
direction \cite{Romero-Isart, Chang, Aspelmeyer-Vitali, Pulsed,
Romero-Isart-Aspelmeyer, Romero-Isart2}. It would thus be very
useful to be able to test the validity of the superposition
principle with the mechanical oscillator initially in {\em thermal
equilibrium} with its environment  (this is our desideratum 1).
Secondly, in many proposals, a high finesse cavity
facilitates the preparation and probing of nonclassical states of
a mechanical object
\cite{Bose-Jacobs-Knight,Marshall,Romero-Isart-Aspelmeyer,
Romero-Isart2} (note, however, an exception \cite{Pepper}).
However, it is demanding for a cavity coupled to a
mechanical resonator to have a high finesse
\cite{diffraction-limit}. Time resolutions in measurements of
cavity fields is also limited (although, see \cite{Pulsed}). Thus
avoiding a cavity will form our desideratum 2.
Matter-wave experiments satisfy desiderata 1
and 2, but require the detection of a spatial interference pattern
generated by an ensemble of particles \cite{Arndt}. Keeping this pattern spatially resolvable and robust to mass
and velocity spreads of the ensemble is challenging -- a challenge
that would probably partially carry over to some optomechanics
\cite{Romero-Isart-Aspelmeyer, Romero-Isart2} and
spin-optomechanics \cite{Duan} schemes.   Avoiding ensembles and
spatially resolved measurements will thus form our desiderata 3
and 4. Lastly, controlling an interference pattern by imparting a
tunable relative phase between superposed paths is the ``litmus
test" of any interference -- yet this has been largely overlooked
in the context of mesoscopic mechanical oscillators. The
controllable relative phase should stem from a field such as
gravity that only interacts with the mass and has no chance to
directly interact with auxillary probes such as the spin. In this
way, by detecting the influence of gravity on the interference of
spin, one can be sure that the mechanical object had been part of
a quantum superposition during an experiment. Proposals so far
have resorted to subtle methodology such as detecting a mass
dependent partial decoherence of a probe
\cite{Bose-Jacobs-Knight,Bose-PRL} or a periodic
decoherence-recoherence dynamics of a  probe
\cite{Bose-Jacobs-Knight,Marshall,
Armour-Blencowe-Schwab,Bose-PRL,Magneto} (this is susceptible to
temperature \cite{Kleckner}). The requirement of a more
transparent interference experiment using a ``controllable
relative phase" between different COM states of a
mechanical oscillator, one which can be solely imparted to the
oscillator, yet can be read from a probe, will thus form our
desideratum 5. We now present a scheme that satisfy all
the 5 desiderata.

Our setup is shown in
Fig.\ref{setup} and consists of a nano-scale diamond bead
containing a single spin-1 NV center levitated by an optical
tweezer in ultra-high vacuum. We need a static magnetic field
gradient to couple the motion of the
bead in the harmonic potential of the tweezer to the
$S=1$ spin of the NV center. This can be generated by a magnetized sphere with a
permanent dipole moment ${\bf m}=(0,0,m_z)$
oriented along the $z$ direction. Let the centers of the
harmonic potential and the magnetized sphere be at $(0,0,0)$ and
$(0,0,z_0)$ respectively. Expanding the magnetic field of the
sphere around $(0,0,0)$, we get
\begin{equation}\label{bxy}
 B_x=-B_0\,x,\;\;B_y=-B_0\,y,\;\;B_z=\frac{\mu_0\,m_z}{2\pi\,|z_0|^3}+2B_0\,z,
\end{equation}
where $B_0=3\mu_0\,m_z/(4\pi z_0^5)$. Therefore  the Hamiltonian describing the interaction between the
spin of the NV center and the vibrational motion can be written
as:
\begin{equation}
 H_{\rm int}=-\lambda\big[2\,S_z\,(c+c^\dag)-\sqrt{\frac{\omega_z}{\omega_x}}\,S_x\,(a+a^\dag)-\sqrt{\frac{\omega_z}{\omega_y}}\,S_y\,(b+b^\dag)\big],
\end{equation}
where
\begin{equation}\label{lambda}
\lambda=\frac{3\mu_0 m_z z_0}{4\pi |z_0|^5}\, g_{NV}\, \mu_B\,
\sqrt{\frac{\hbar}{2\, m \,\omega_z}},
\end{equation}
$m$ being the mass of the bead, $g_{NV}$ the Land\'e factor of the
NV center and $\mu_B$ the Bohr magneton. To this we must add the
free Hamiltonian of the NV center and of the bead
motion, i.e. $H_{\rm free}=D\,S_z^2+\hbar(\omega_xa^\dag
a+\omega_yb^\dag b+\omega_zc^\dag c)$. The anisotropy $D S_z^2$ of the NV
center is not a problem as the
possibility to align the $z$ axis of the defect in the
nano-diamond to any desired direction has been recently
demonstrated \cite{alignment, nano-ESR1}. We consider the Zeeman splitting of  $\left|+1\right>$
and $\left|-1\right>$ due to the zeroth order expansion of $B_z$ to be cancelled by addition of a
uniform magnetic field along $z$.

\begin{figure}
\begin{center}
\includegraphics[width=3.5in, clip]{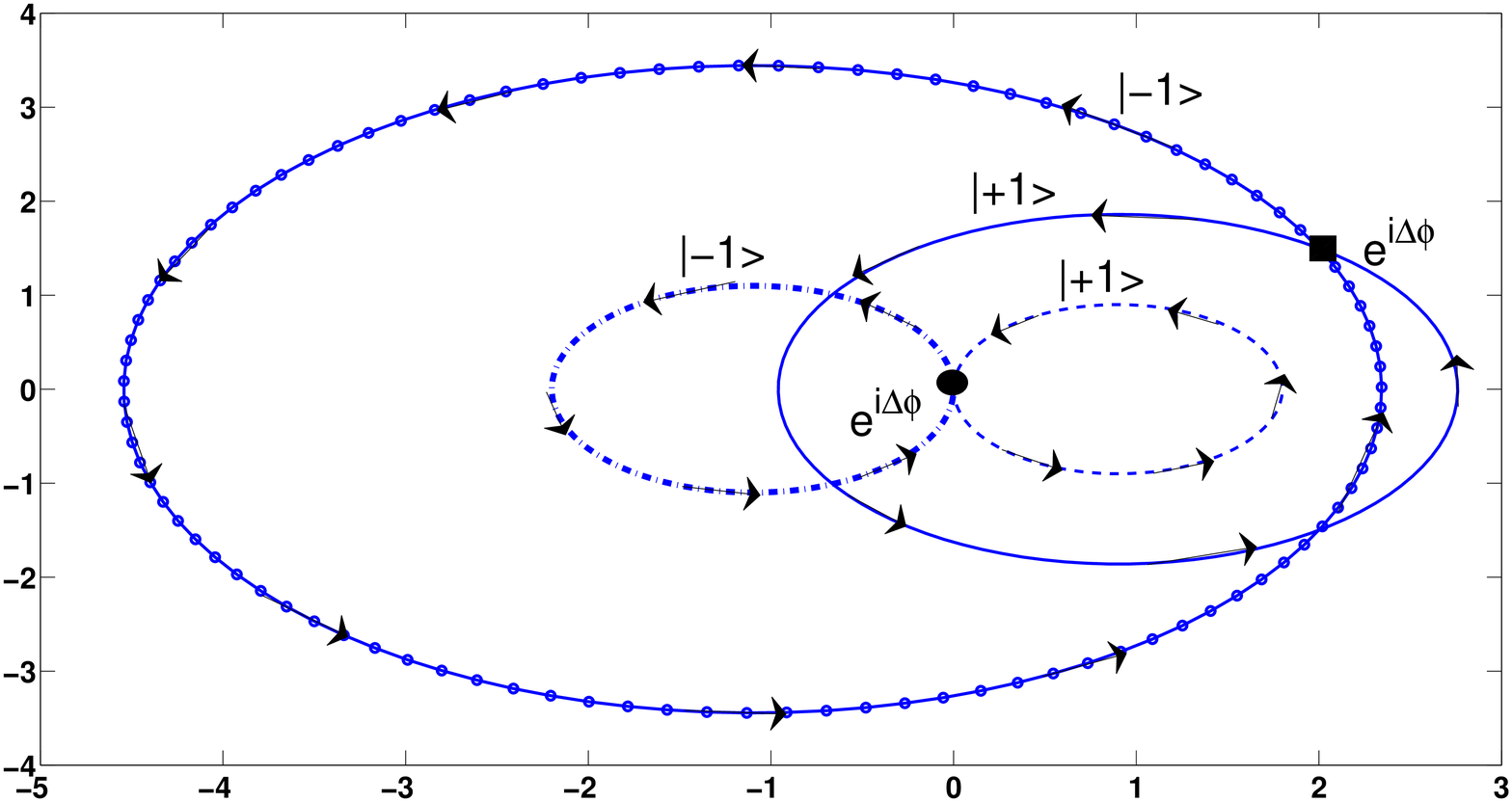}
 \caption{The phase space evolution of two arbitrary initial coherent states $\beta=0$ (filled circle) and $\beta=2+i1.5$ (filled square)
 with $\lambda=0.5 \hbar \omega_z$ and $\Delta \lambda=0.05 \hbar \omega_z$.  The arrows show the direction of evolution of the coherent states with
 increasing time for a whole time period $t_0=2\pi/\omega_z$. The return of each component to the initial coherent state at $t_0$ is evident.
 A relative phase
 $e^{i\Delta \phi_{\text{Grav}}}$ appears between the $|+1\rangle$ and $|-1\rangle$ component due to the gravitational potential difference between
 the centers of the phase space trajectories for $|-1\rangle$ (more towards the left of the figure) and $|+1\rangle$
 (more towards the right of the figure).}
\label{phase-space}
\end{center}
\end{figure}

Finally, we assume  $\omega_x,\omega_y\gg\omega_z$ and
add the interaction with the Earth's gravitational field
$mgz\cos\theta$, where $\theta$ is the angle between the $z$
direction and the free fall acceleration (see Fig. 1). Therefore
we obtain the Hamiltonian:
\begin{equation}\label{Ham}
 H=D\,S_z^2+\hbar\omega_zc^\dag c-2(\lambda\,S_z+\Delta\lambda)(c+c^\dag),
\end{equation}
with
\begin{equation}\label{deltalambda}
 \Delta\lambda=\frac{1}{2}mg\cos\theta\sqrt{\frac{\hbar}{2m\omega_z}}.
\end{equation}
The Hamiltonian above represents a
harmonic oscillator whose center depends on the eigenvalue of $S_z$. 
For each of $S_z=-1,\,0,$ and $+1$, we can
calculate the evolution of the oscillator, when its initial
state is a coherent state $\left|\beta\right>$ of the
harmonic well centered at $z=0$. Transforming by means of an
appropriate displacement operator, we can calculate the evolution
governed by a new Hamiltonian corresponding to a free harmonic
oscillator with no displacement.
Transforming back to the original representation, we
get the three spin-dependent evolutions
\begin{equation}
 \left|\beta\right>\left|s_z\right> \rightarrow \left|\beta (t, s_z)\right>\left|s_z\right>
\label{evolution}
\end{equation}
where, denoting $u=2(s_z\,\lambda-\Delta\lambda)/\hbar\omega_z$,
\begin{eqnarray}
\left|\beta (t, s_z)\right> =&
\mathrm{e}^{-\frac{i}{\hbar}(D-\hbar\omega_z u^2)t}
 \mathrm{e}^{iu^2\sin(\omega_z t)}\nonumber\\
 &
\times\left|(\beta-u)\mathrm{e}^{-i\omega_z t}+u\right>,
\label{dynamics}
\end{eqnarray}
with the ket on the right hand side being a coherent state and $s_z$ being an eigenvalue of $S_z$.
%
%
%
The above time evolution is illustrated in phase space for two
distinct values of $\beta$ in Fig.\ref{phase-space}. A striking
feature of this evolution is
that at time $t_0=2\pi/\omega_z$ the oscillator state returns to
its original coherent state $\beta$, for {\em any} $\beta$ and
$s_z$. This feature implies that spin measurements at $t_0$ will
be unaffected by any randomness in the initial motional
state of the oscillator. We now exploit this key feature to
propose a Ramsey interferometric scheme that does not require any
preparation of the initial COM state, yet detects a
phase difference stemming from superpositions involving distinct COM positions.

{\em Ramsey interferometry -} Consider that the system
starts from a separable state
$\left|\beta\right>\left|s_z=0\right>$. The first step applies a microwave
pulse corresponding to the Hamiltonian
$H_{mw}=\hbar\Omega\left(\left|+1\right>\left<0\right|+\left|-1\right>\left<0\right|+\mathrm{h.c.}\right)$,
with $\Omega$ much larger than any other coupling constant, so
that we can neglect any other interaction when applying the pulse.
With the pulse duration $t_p=\pi/(2\sqrt{2}\Omega)$, the
spin state becomes a superposition with equal amplitudes of
$\left|+1\right>$ and $\left|-1\right>$:
$ \left|\Psi(0)\right>=\left|\beta\right>\left(\frac{\left|+1\right>+\left|-1\right>}{\sqrt{2}}\right)$,
which we take as the initial state for the interaction under the Hamiltonian (\ref{Ham}).
At the interaction time $t$, the state is then
\begin{equation}
 \left|\Psi(t)\right>=\left(\frac{\left|\beta(t,+1)\right>\left|+1\right>+\left|\beta(t,-1)\right>\left|-1\right>}{\sqrt{2}}\right),
\end{equation}
which is the superposition we intend to evidence. Note from the
expressions of $\left|\beta(t,\pm 1)\right>$ in
Eq.(\ref{dynamics}) that separated coherent states are involved in
the above superposition along with phases due to gravitational
potential, which are finally going to evidence the above
superposition. The state after an oscillation period
$t_0=2\pi/\omega_z$ is
\begin{equation}
 \left|\Psi(t_0)\right>=\left|\beta\right>\left(\frac{\left|+1\right>+\mathrm{e}^{i\Delta\phi_{\text{Grav}}}\left|-1\right>}{\sqrt{2}}\right),
\end{equation}
where, after dropping a global phase factor, we have
\begin{equation}\label{deltaphi}
 \Delta\phi_{\text{Grav}}=\frac{16\lambda\,\Delta\lambda}{\hbar^2\omega_z}\,t_0.
\end{equation}
To reveal $\Delta\phi_{\text{Grav}}$ we apply
$H_{mw}$ again. After a time $t_p$, the population of the spin state with $S_z=0$ is:
\begin{equation}\label{population}
 P_0(t=t_0+t_p)=\cos^2\left(\frac{\Delta\phi_{\text{Grav}}}{2}\right),
\end{equation}
which gives a direct connection between the value of the phase
shift and spin population. As
$\Delta\phi_{\text{Grav}} \propto g$ (gravitational acceleration)
can never appear as a relative phase between spin states unless
spatially separated states of the COM were involved in
the superposition
$\left(\left|\beta(t,+1)\right>\left|+1\right>+\left|\beta(t,-1)\right>\left|-1\right>\right)/\sqrt{2}$
for $0< t < t_0$, the detection of
$\Delta\phi_{\text{Grav}}$ evidences such a superposition.


{\em Thermal effects -} We can exploit the facts that the results
given above are independent of the amplitude $\beta$ (see
Fig.\ref{phase-space} for an illustration) and that any thermal
state $\rho_{th}$ of the motion can be written as $\rho_{th}= \int
\mathrm{d}^2\beta\,P_{th}(\beta)\left|\beta\right>\left<\beta\right|$, where $P_{th}$ is the
Glauber $P$ representation for the thermal state.
Considering the
following initial state for the composite system:
\begin{equation}
 \rho(t=0)=\rho_{th}\otimes\frac{1}{2}\left(\left|+1\right>+\left|-1\right>\right)\left(\left<+1\right|+\left<-1\right|\right),
\end{equation}
at time $t_0$ we have
\begin{eqnarray}
 \rho(t_0)&=&\int
 \mathrm{d}^2\beta\,P_{th}(\beta)\left|\beta\right>\left<\beta\right|\\
 &\otimes&\frac{1}{2}\left(\left|+1\right>+
 \mathrm{e}^{i\Delta\phi_{\text{Grav}}}\left|-1\right>\right)\left(\left<+1\right|+
 \mathrm{e}^{-i\Delta\phi_{\text{Grav}}}\left<-1\right|\right),\nonumber
\end{eqnarray}
which shows that, after one period, the state of the system is
again factorizable and that the phase difference accumulated by
the spin states is not affected by the thermal motion. Basically, though a mixture of several
Schr\"odinger cats
$\left|\beta(t,+1)\right>\left|+1\right>+\left|\beta(t,-1)\right>\left|-1\right>$
is generated for $0< t < t_0$, the interference between the
components $\left|\beta(t,+1)\right>\left|+1\right>$ and
$\left|\beta(t,-1)\right>\left|-1\right>$ of the cat is
independent of $\beta$. This immunity of the interference to
thermal states hinges on the mass being trapped in a
harmonic potential. We assume that anharmonic effects of the
trapping potential will be avoided by
feedback cooling of our oscillator to mK temperatures
\cite{Raizen,Novotny}. We will also justify that the heating {\em during} evolution is negligible for $0\leq t \leq t_0$.

{\em Experimental parameters - }We now give the
parameters necessary to obtain a good visibility of
the interferometric fringes in a setup in which we
are allowed to vary the angle $\theta$. As realistic values, we
consider $\omega_z \sim 100$ kHz and diamond spheres whose radius
$R \sim 100$ nm, so that, considering the density
$3000\,\mathrm{kg}/\mathrm{m}^3$ for diamond, the
corresponding mass is $\sim 1.25 \times 10^{-17}$ kg. A
good visibility of interferometry fringes in the population in Eq.
(\ref{population}) is given for
$K=8\lambda\,\Delta\lambda\,t_0/(\hbar^2\omega_z\cos\theta)\sim10$,
which makes the value of the population change completely from $0$
to $1$ when $\theta$ varies between $\pi/2-\pi/20$ and $\pi/2$
(the $z$ axis is horizontal for $\theta=\pi/2$). Assuming that the
magnetic field in Eq. (\ref{bxy}) is generated by a magnetized
sphere with radius $r_0=40\,\mu$m and magnetization $M=1.5 \times
10^{6}\, \mathrm{A}/\mathrm{m}$ (typical for commercial
magnets), and $z_0=120\,\mu$m, we get, according to $m_z=M\cdot
(4\pi/3)r_0^3$ and Eqs. (\ref{lambda}) and (\ref{deltalambda}),
the desired value of $K$.

Corresponding to such values, it is worth noting that the maximum
separation between the coherent states involved in the
interferometric scheme is, according to Eq. (\ref{evolution}),
$4\lambda/\hbar\omega_z\simeq 0.03$, which shows that, during the
scheme proposed, the superpositions involved are not macroscopic.
This can be advantageous since it allows to have very small values
for the motional decoherence rates, whose maximum value will be
$\gamma_{sc}\cdot |2\lambda/\hbar\omega_z|^2$, where $\gamma_{sc}$
is the decay rate associated with photon scattering from the trapping
laser \cite{Chang}. For the diamonds used in the laboratory, with
a dielectric permittivity $\epsilon=1.5$, we have
$\gamma_{sc}/\omega_z=(16\pi^3/15)((\epsilon-1)/(\epsilon+2))(R^3/\lambda_0^3)\simeq
5\times 10^{-3}$, for spheres of radius $R=100$ nm  and trapping wavelength $\lambda_0 \sim 1\, \mu$m. Thus the values we found allow us to neglect the
decoherence due to light scattering for the duration $t_0 \sim 10
\mu$s of our scheme. For the same reason heating, in the form of
random momentum kicks {\em during} the $t_0 \sim 10 \mu$s
duration, can be neglected. Moreover, feedback cooling to mK
temperatures, which has already been demonstrated \cite{Raizen, Novotny}, makes the thermal state phonon
number of the levitated diamond to be $\sim 1000$, which is well
below the the energy before which the harmonic approximation
starts to break down. Thus
our protocol should work even under a thermal environment.

The detrimental effects on the scheme are therefore only due to
the dephasing of the NV center due to other spins in the diamond
lattice. The value for this time ($T_2$) in bulk diamond is
exceptional ($\sim$ ms \cite{diamond-coherence}). While such a
value would pose no dephasing in the timescale of our scheme, for
$R \sim 20$ nm  nano-diamonds $T_2\sim$ $10\,\mu$s is
significantly lower due to interactions between the spin and
defects on the surface \cite{nano-ESR2}. Our larger ($100$ nm)
beads would remove the surface further from the NV center and
improve the coherence. Additionally, we propose to use to spin
echo techniques along with a sudden change of the orientation
(angle $\theta$) of the experimental setup, which is now
experimentally feasible \cite{bishop}. If after one period $t_0$
we only apply a mw pulse so that the spin state
$\left|+1\right\rangle$ goes to the state $\left|-1\right\rangle$
and vice versa, at time $t=2t_0$ the random phases (at the origin
of dephasing) acquired by the spin states during the second
oscillation will cancel the ones acquired during the first
oscillation, minimizing the dephasing effects. However, the spin
echo pulse would also cause the spin states to acquire opposite
values for the gravitational phases during the second period,
cancelling out the phases accumulated after the first oscillation.
This problem persists even if the refocusing pulse is orthogonal
to the preparation pulse as in the CPMG experiment, where the
sequence of pulses
$(\pi/2)_x-[\pi_y-\mathrm{echo}]_\mathrm{repeat}$ is used
\cite{nano-ESR2}. To avoid this, it is enough to reverse the
direction of the $z$ axis with respect to the horizontal plane, so
that $\theta\rightarrow\pi-\theta$, which corresponds to having
the same evolutions as in Eq. (\ref{evolution}) with the
substitution $\Delta\lambda\rightarrow-\Delta\lambda$:
consequently, at time $t=2t_0$, the phase difference acquired
between $\left|-1\right>$ and $\left|+1\right>$ will be twice the
one given in Eq. (\ref{deltaphi}). The detection scheme can then
proceed as described before, with an improved visibility. For our
parameters \cite{nano-ESR2},  $t_0/T_2\sim 1$, we still expect a
visible fringe, though its contrast is reduced to $1/e \sim 0.36$.

Our proposal combines desirable aspects in a feasible experiment
while some of individual elements have been studied before. For
example, the ability to map the decoherence of a mesoscopic
oscillator to a probe field \cite{Bose-Jacobs-Knight} or qubit
\cite{Bose-PRL} has been appreciated before, as well as its
possibility for a thermal state. In Ref. \cite{vacanti}, the
mapping of a geometric phase to a probe qubit has been studied --
here we are exploiting a similar idea, but, instead, mapping a
dynamical phase from a gravitational potential term in the
Hamiltonian. The dynamics entailed by Eqs.(\ref{dynamics}) has
been exploited in several papers
\cite{Bose-Jacobs-Knight,Marshall,Armour-Blencowe-Schwab,Duan,Bose-PRL,Magneto},
though not for Ramsey interferometry. Similarly, levitated
diamonds containing an NV center have very recently been
considered for generating pure Schr\"odinger cats \cite{Duan} (in
fact, while the work of this paper was in progress). We differ
from this scheme in terms of the absence of a cavity and the
gravitationally induced Ramsey interferometry. Thus our scheme is
easy to probe without having to read out the motional state.
Indeed this latter aspect is the central advance of our scheme
over previous works as it allows a spin probe (which can be
measured fast) to accumulate a gravitational phase shift. Perhaps
other applications of the qubit-oscillator coupling
$S_z\,(c+c^\dag)$ \cite{vacanti,apps} can also be studied with our
setup.

Macroscopic limits of quantum superpositions can be reached in two
ways -- larger masses and larger spatial separations between the
superposed components. Here we will be able to test quantum
superpositions for a large mass ($\sim 10^{-17}$ kg), but the
physical separation between the COM parts of the superposed states
$\left|\beta(t,+1)\right>\left|+1\right>$ and
$\left|\beta(t,-1)\right>\left|-1\right>$ is quite small (distance
$\sim 0.03$ in phase space which is about $0.15$ picometers; it
can be increased possibly to about $\sim 3$ in phase space by
decreasing $z_0$ to $1 \mu$m and $r_0 \sim 1/3 \mu$m by using
nano-magnets).  However, that $\Delta\phi_{\text{Grav}}$ is still
accumulated and significant to give a full fringe
($P_0(t=t_0+t_p)$ ranging from $0$ to $1$ as $\theta$ is varied)
is interesting. Though in the end we only measure the spin, it
still evidences the involvement of the COM in superpositions of
the hybrid system as it is impossible for a spin to obtain a
gravitational phase shift on its own. Our proposal satisfies a few
desiderata (modest cooling, no cavities, ensembles or spatially
resolved measurements, Ramsey interferometry with controllable
phase) which potentially makes the path to extend quantum
mechanics to more macroscopic objects simpler.

We acknowledge the EPSRC grant EP/J014664/1 with which MS is
funded. GWM is supported by the Royal Society. MSK acknowledges
support of the Qatar National Research Fund (Grant NPRP
4-554-D84). We are very thankful to Hendrik Ulbricht and Andreas
Schell for comments which have improved the manuscript.


\begin{thebibliography}{10}
\bibitem{Leggett}
A. O. Caldeira and A. J. Leggett, Annals of Physics {\bf 149}, 374 (1983).
\bibitem{Zurek}
W. H. Zurek, Phys. Today {\bf 44} (10): 36, (1991).
\bibitem{GRWP}
G. C. Ghirardi, A. Rimini, and T. Weber, Phys. Rev. D {\bf 34}, 470 (1986); P. Pearle, Phys. Rev. A39, 2277 (1989).
\bibitem{Penrose}
R. Penrose, General Relativity and Gravitation, {\bf 28}, 581 (1996).
\bibitem{RMP}
A. Bassi, K. Lochan, S. Satin, T. P. Singh, and H. Ulbricht, Rev. Mod. Phys. {\bf 85}, 471 (2013).
\bibitem{Leggett-Garg}
A. J. Leggett, A. Garg, Phys. Rev. Lett. {\bf 54}, 857 (1985).
\bibitem{flux-qubits}
J. R. Friedman, V. Patel, W. Chen, S. K. Tolpygo \& J. E. Lukens, Nature {\bf 406}, 43 (2000).
\bibitem{Arndt}
S. Gerlich, S. Eibenberger, M. Tomandl, S. Nimmrichter, K.
Hornberger, P. J. Fagan, J. T\"{u}xen, M. Mayor, and M. Arndt,
Nature Comm. {\bf 2}, 263 (2011).
\bibitem{Bose-Jacobs-Knight}
S. Bose, K. Jacobs, and P. L. Knight, Phys. Rev. A {\bf 59}, 3204 (1999).
\bibitem{Marshall}
W. Marshall, C. Simon, R. Penrose, and D. Bouwmeester, Phys. Rev. Lett., {\bf 91} 130401 (2003).
\bibitem{Armour-Blencowe-Schwab}
A.D. Armour, M.P. Blencowe, and K. Schwab, Phys. Rev. Lett. {\bf 88},  148301 (2002).
\bibitem{Rabl}
P. Rabl, P. Cappellaro, G. Dutt, L. Jiang, J. R. Maze, and M. D. Lukin, Phys. Rev. B {\bf 79}, 041302 (2009).
\bibitem{Thew} N. Bruno, A. Martin, P. Sekatski, N. Sangouard, R. Thew, and N.
Gisin, arXiv:1212.3710 [quant-ph].
\bibitem{Barker}
P. F. Barker and M. N. Schneider, Phys. Rev. A {\bf 81}, 023826 (2010).
\bibitem{Romero-Isart}
O. Romero-Isart, M. L. Juan, R. Quidant, and J. I. Cirac, New J. Phys. {\bf 12}, 033015 (2010).
\bibitem{Chang}
D. E. Chang, C. A. Regal, S. B. Papp, D. J. Wilson, J. Ye, O. Painter, H. J. Kimble and P. Zoller, Proc. Natl. Acad. Sci. USA {\bf 107}, 1005 (2010).
\bibitem{Aspelmeyer-Vitali}
M. Paternostro, D. Vitali, S. Gigan, M. S. Kim, C. Brukner, J. Eisert, M. Aspelmeyer, Phys. Rev. Lett. {\bf 99}, 250401 (2007).
\bibitem{Pulsed}
M. R. Vanner, I. Pikovski, G. D. Cole, M. S. Kim, C. Brukner, K. Hammerer, G. J. Milburn, M. Aspelmeyer, Proc Natl Acad Sci USA {\bf 108}, 16182 (2011).
\bibitem{Romero-Isart-Aspelmeyer}
O. Romero-Isart, A. C. Panzer, F. Blaser, R. Kaltenbaek, N. Kiesel, M. Aspelmeyer, and J. I. Cirac, Phys. Rev. Lett.
{\bf 107}, 020405 (2011).
\bibitem{Romero-Isart2}
O. Romero-Isart, Phys. Rev. A {\bf 84}, 052121 (2011).
\bibitem{Pepper}
B. Pepper, R. Ghobadi, E. Jeffrey, C. Simon, D. Bouwmeester, Phys. Rev. Lett. {\bf 109}, 023601 (2012).
\bibitem{diffraction-limit}
D. Kleckner, W. T. M. Irvine, S. S.
R. Oemrawsingh and D. Bouwmeester, Physical Review A 81, 043814 (2010).
\bibitem{Duan}
Zhang-qi Yin, Tongcang Li, Xiang Zhang, L. M. Duan, arXiv:1305.1701.
\bibitem{Bose-PRL}
S. Bose, Phys. Rev. Lett. {\bf 96}, 060402 (2006).
\bibitem{Magneto}
O. Romero-Isart, L. Clemente, C. Navau, A. Sanchez, J. I. Cirac, Phys. Rev. Lett. {\bf 109}, 147205 (2012).
\bibitem{Kleckner} D. Kleckner, I. Pikovski, E. Jeffrey, L. Ament, E. Eliel, J. van den Brink and D. Bouwmeester,
New Journal of Physics {\bf 10}, 095020 (2008).



\bibitem{alignment} M. Geiselmann, M. L. Juan, J. Renger, J. M. Say, L. J. Brown, F. J. Garc\'ia de Abajo, F. Koppens, and R. Quidant,
Nature Nanotechnology {\bf 8}, 175 (2013).

\bibitem{nano-ESR1}
V. R. Horowitz, B. J. Alemán, D. J. Christle, A. N. Cleland, and D. D. Awschalom, Proc. Natl. Acad. Sci. USA  {\bf 109}, 13493 (2012).

\bibitem{nano-ESR2}
A. Laraoui, J. S. Hodges, and C. A. Meriles, Nano Lett. {\bf 12}, 3477 (2012).


\bibitem{Raizen}
T. Li, S. Kheifets, and M. G. Raizen, Nat. Phys. {\bf 7}, 527-530
(2011).

\bibitem{Novotny}
J. Gieseler, B. Deutsch, R. Quidant, and L. Novotny,
Phys. Rev. Lett. {\bf 109}, 103603 (2012).

\bibitem{diamond-coherence}B. Naydenov, F. Dolde, L. T. Hall, C. Shin, H. Fedder, L. C. L. Hollenberg, F. Jelezko, and J.
Wrachtrup, Phys. Rev. B {\bf 83}, 081201 (2011) ; K. D. Jahnke, B.
Naydenov, T. Teraji, S. Koizumi, T. Umeda, J. Isoya, and F.
Jelezko Applied Physics Letters {\bf 101}, 012405 (2012); P. L.
Stanwix, L. M. Pham, J. R. Maze, D. Le Sage, T. K. Yeung, P.
Cappellaro, P. R. Hemmer,A. Yacoby, M. D. Lukin, and R. L.
Walsworth, Phys. Rev. B 82, 201201(R) (2010); P. C. Maurer, G. Kucsko, C. Latta, L. Jiang, N. Y. Yao, S.
D. Bennett, F. Pastawski, D. Hunger, N. Chisholm, M. Markham, D.
J. Twitchen, J. I. Cirac, and M. D. Lukin, Science {\bf 336}, 1283
(2012);  H. Bernien, B. Hensen, W. Pfaff, G. Koolstra, M. S. Blok,
L. Robledo, T. H. Taminiau, M. Markham, D. J. Twitchen, L.
Childress, and R. Hanson, Nature {\bf 497}, 86 (2013); M. V.
Gurudev Dutt, L. Childress, L. Jiang, E. Togan, J. Maze, F.
Jelezko, A. S. Zibrov, P. R. Hemmer, M. D. Lukin, Science {\bf
316}, 1312 (2007).



\bibitem{bishop} A. I. Bishop and P. F. Barker, Rev. Sci. Instrum. 77 , 044701
(2006).

\bibitem{vacanti} G. Vacanti, R. Fazio, M. S. Kim, G. M. Palma, M. Paternostro, and V.
Vedral, Phys. Rev. A {\bf 85}, 022129 (2012).

\bibitem{apps} T. P. Spiller, K. Nemoto, S. L. Braunstein, W. J. Munro, P. van
Loock, G. J. Milburn New J. Phys. {\bf 8}, 30 (2006); T. A. Brun,
Hsi-Sheng Goan, Phys. Rev. A {\bf 68}, 032301 (2003); L. Garcia,
R. W. Chhajlany, Y. Li, Lian-Ao Wu, arXiv:1305.0290; Y. Li, L. Wu,
Y. Wang, and L. Yang, Phys. Rev. B {\bf 84}, 094502 (2011).



\end{thebibliography}
\end{document}